\documentclass{ws-procs9x6}

\begin{document}

\title{
PROBING IN-MEDIUM VECTOR MESONS BY DILEPTONS \\
AT HEAVY-ION COLLIDERS\footnote{
Lecture given at the 42-nd Predeal International Summer School in Nuclear Physics
{\it Collective Motion and Phase Transitions in Nuclear Systems},
28 August - 9 September, Predeal, Romania.
}
}

\author{M. I. KRIVORUCHENKO}

\address{Institute for Theoretical and Experimental Physics, B. Cheremushkinskaya 25\\
117259 Moscow, Russia\\
E-mail: mikhail.krivoruchenko@itep.ru\\
}

\begin{abstract}
An introduction to physics of in-medium hadrons 
with special emphasis towards modification of vector meson properties in 
dense nuclear matter is given. We start from remarkable analogy between
the in-medium behavior of atoms in gases and hadrons in nuclear matter. 
Modifications of vector meson widths and masses can be registered experimentally 
in heavy-ion collisions by detecting dilepton spectra from decays of 
nucleon resonances and light unflavored mesons including $\rho$- and $\omega$-mesons. 
Theoretical schemes 
for description of the in-medium hadrons are reviewed and recent experimental 
results of the NA60 and HADES collaborations on the dilepton production 
are discussed. 
\end{abstract}


\bodymatter

\section{Introduction}\label{aba:sec1}
Elementary particles as ground-state excitations are characterized differently 
in the vacuum and in the dense matter. This effect is known from the solid state physics 
where elementary excitations are traditionally called {\it quasi-particles} to indicate 
medium-induced modifications. The medium effects manifest themselves in a wide range of
physical phenomena ranging from shifts and broadening of energy levels of pionic atoms to 
matter-affected neutrino oscillations.

In-medium modifications of hadron properties are of special interest in order to 
test QCD at finite densities and temperatures and, in particular, for better
understanding of the chiral and quark-hadron phase transitions and QCD
confinement problem. Heavy-ion collisions provide a unique possibility to
create nuclear matter under extreme conditions and to study hadron
physics around phase transitions points. One of the best direct
probes to measure in-medium masses and widths of vector mesons are dileptons
$e^{+}e^{-}$ or $\mu^{+}\mu^{-}$ which, being produced, leave the reaction zone essentially undistorted by final-state interactions. 

The key problem is the dependence of elementary excitations on density and temperature. 
The knowledge of such dependence allows to use elementary particles as probes to measure
thermodynamical properties of highly compressed 
nuclear mater, improving thereby our knowledge of QCD, nuclear physics and the phenomenology 
of strong interactions.

In this lecture we discuss main effects leading to modifications of in-medium spectral functions of hadrons. In the next Sect., a remarkable analogy between atoms in gases and hadrons in nuclear matter
is discussed. In Sect. 3, we show how the concept of collision broadening of particles, originated
from the atomic physics, applies to behavior of nucleon resonances in nuclear matter. Sects. 4 and 5 provide a summary, respectively, of theoretical and experimental works on in-medium modifications of vector mesons and dilepton production. Sect. 6 deals with elementary sources of dileptons. The extended vector meson dominance model (eVMD) is formulated in Sect. 7 and its gauge invariance is proved in Sect. 8. Sect. 9 discusses the recent experimental results (summer 2006) from the NA60 and HADES collaborations. 

\section{Atoms in gases}

The concept of collision broadening is discussed in atomic physics 
since last century. The physics behind the modifications of 
atomic spectral lines in gases represents the obvious theoretical interest, 
because it has significant features in common with physics of behavior of 
hadrons in nuclear matter.

\subsection{Radiation damping}

It is known that radiative damping results to finite widths of atomic spectral 
lines. The energy deposited during transition to a lower energy state can never 
be monochromatic, but distributed according to the Lorentz formula
\begin{equation}
dI(\omega) = \frac{I}{2\pi} \frac{\Gamma d\omega}{(\omega - \omega_{0})^2 + (\Gamma/2)^2} 
\label{LORE}
\end{equation}
where 
\begin{equation}
\Gamma = \Gamma_{a} + \Gamma_{b}, 
\label{GAMMA}
\end{equation}
with $\Gamma_{a}$ and $\Gamma_{b}$ being the vacuum widths of levels $a$ and $b$, 
respectively, and $\omega_{0} = E_{a} - E_{b}$ is the transition energy. 

Recall its quantum-mechanical derivation: The wave function of state $a$ equals
$\Psi \sim \exp(-iE_{a}t - \frac{1}{2}\Gamma_{a}t)$, similarly for $b$.
Transition amplitude $a \to b$ + photon of energy $\omega$ is given by 
$A_{ba} \sim (\exp(-i\omega t)\Psi_{b})^{*}\Psi_{a}$. Evaluation of the integral
$\int_{- \infty}^{+ \infty} |A_{ba}|^2 dt$ gives spectral density (\ref{LORE}).

Radiation damping is inherent to any radiating system. It results to broadening of spectral 
lines of isolated atoms. Thermal motion of atoms and collisions of particles affect the 
profile of lines also.

\subsection{Doppler effect}

Doppler effect results to an additional broadening of spectral lines due to 
motion of atoms in gases. It has a purely kinematic origin.

\subsection{Collision broadening}

Lorentz attributed collision broadening of spectral lines to a decoherence effect accompanied scattering of a probing atom with surrounding atoms and electrons (see e.g. \cite{SOBEL}, Chap. 10). Consider an excited atom which experiences at random times $t_{i}$ collisions with 
surrounding particles. Its emission intensity is summed up coherently between two collisions
for $\tau \in [t_{i},t_{i + 1}]$ and decoherently with respect to time intervals 
$[t_{i},t_{i + 1}]$. One has
\begin{equation}
dI(\omega) \sim \frac{d\omega}{2\pi N}\sum_{i = 1}^{N}\left|\int_{t_{i}}^{t_{i + 1}}
{
dt \exp(-i(\omega - \omega_{0})t - \frac{\Gamma}{2}t)
}\right|^2.
\label{LOOO}
\end{equation}
One has to assume further that waiting times for sequential collisions $\tau_{i} = t_{i + 1} - t_{i}$ are distributed exponentially
\begin{equation}
dW(\tau) = \frac{d\tau}{\tau_{0}}\exp(-\tau/\tau_{0}).
\end{equation}
Such a hypothesis is equivalent to the requirement of the Poisson distribution of 
surrounding uncorrelated particles, $\tau_{0}$ has the meaning of collision time.
Replacing the average over collisions $\frac{1}{N}\sum_{i = 1}^{N}\ldots $ 
with the integral $\int dW(\tau) \ldots $, one arrives at the Lorentz formula 
(\ref{LORE}), with $\Gamma$ replaced by 
$\Gamma^{total} = \Gamma + 2/\tau_{0}$.

According to Eq.(\ref{GAMMA}), the radiation width $\Gamma$ represents the sum of 
radiation widths of levels $a$ and $b$, so one has to split $2/\tau_{0}$ among two levels:
\begin{equation}
\Gamma_{a}^{total} = \Gamma_{a} + 1/\tau_{0}
\label{THREE}
\end{equation}
and similarly for $b$. The value 
\begin{equation}
\Gamma_{coll} = 1/\tau_{0}
\label{coll}
\end{equation}
is called {\it collision width}. 

We wish to bring attention to two circumstances inherent to the above 
approach: 
\begin{itemize}
\item Collision broadening is physically attributed to decoherence. 
\item The coefficient at $1/\tau_{0}$ in Eq.(\ref{coll}). 
\end{itemize}

The modern schemes use field-theoretic methods which do not refer to decoherence, 
although predict additional broadening inverse proportional to $\tau_{0}$
in the exact agreement with the Lorentz theory. The interplay between the decoherence assumption 
and field-theoretic methods is interesting. The collision width is determined as an imaginary part of (obviously coherent) sum of forward scattering amplitudes with surrounding particles. Due to the quantum-mechanical optical theorem, it is equal to noncoherent sum of cross sections with surrounding particles. The coefficient at $1/\tau_{0}$ appears to be exact, although Eq.(\ref{coll}) is derived using assumptions of the {\it instantaneous} interactions and the {\it decoherence} which are not immediately evident. In the first order to the density, however, there are no corrections to Eq.(\ref{coll}) provided cross sections are calculated quantum-mechanically. The results of the Lorentz theory are therefore recovered using the field-theoretic methods.

\begin{figure}[!htb]
\vspace{-5mm}
\begin{center}
\includegraphics[angle=0,width=6cm]{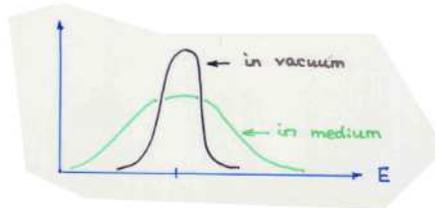}
\vspace{-8mm} 
\caption{Schematic representation of broadening of atomic energy levels 
in gases.}
\label{sig1_fig}
\end{center}
\end{figure}
\vspace{-5mm}

It is instructive to follow an elementary approach of Ref. \cite{Kondratyuk:1994ah}.

The free path length $\ell_{f}$ of a hydrogen atom (or any other particle) in a gas with 
density $n$ is determined by its cross section $\sigma$ for scattering with other particles 
(atoms, molecules, ions, and electrons). There should be one particle inside of a cylinder 
of height $\ell_{f}$ and base $\sigma$. This condition gives
\begin{equation}
\ell_{f} = \frac{1}{n\sigma}.
\label{lfpath}
\end{equation}

The disappearance of atoms from atomic beam as a function of distance $\ell$ is described by an exponential law. Let a hydrogen atom in a state $a$ moves with velocity $v$, then $\ell = vt$. One finds
\begin{equation}
N_{a}(\ell) = N_{a}(0)\exp(-\ell/\ell_{f})~~\&~~ \ell = vt~~\Rightarrow ~~N_{a}(t) = N_{a}(0)\exp(-n \sigma v t). 
\nonumber
\end{equation}

\vspace{-3mm}
\begin{figure}[!htb]
\begin{center}
\includegraphics[angle=0,width=9cm]{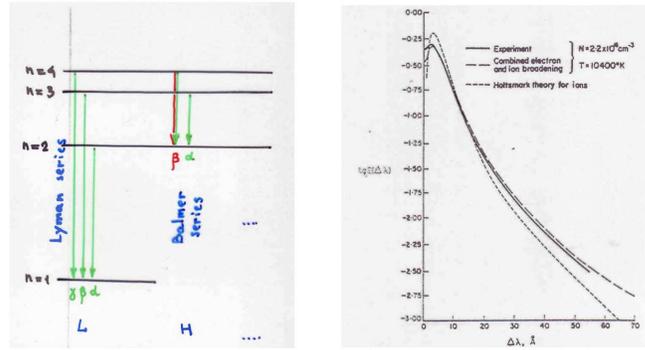}
\caption{{\bf Left panel}: Schematic representation of hydrogen 
atom energy levels and the associated Lyman (L) and Balmer (H) series $\alpha, \beta, \gamma, \ldots$.
{\bf Right panel}: Logarithm of the spectral density of the $H_{\beta}$ line of hydrogen-like atom 
HeI at density of $n = 2.2\times 10^{16}$ cm$^{-3}$ and temperature of $T = 10400$ K 
as a function of the wavelength shift $\Delta \lambda = \lambda - \lambda_{0}$ where 
$\lambda_{0} = 1/\omega_{0}$. The solid line is the experiment. The dashed lines show 
theoretical predictions. 
The spectral density is broadened by a few $\AA$ and has two peaks at 
$\Delta \lambda \sim $ $\pm$ a few $\AA$
(from Ref.\cite{BORGE}).
}
\label{sig1_figu}
\end{center}
\end{figure}
\vspace{-6mm}

In quantum mechanics, decays of a quasistationary state $a$ are governed by equation 
\begin{equation}
|\Psi_{a}(t)|^{2} = |\Psi_{a}(0)|^{2} \exp(-\Gamma_{a} t). \label{der}
\end{equation}
This equation looks like the one describing the disappearance of 
particles from beam due to collisions. Since two distinct mechanisms exists 
for the disappearance and $N_{a}(t) \sim |\Psi_{a}(t)|^{2}$, the right idea 
is to combine two widths:
\begin{equation}
\Gamma^{total}_{a} = \Gamma_{a} + \Gamma_{coll}
\label{tot}
\end{equation}
where $\Gamma_{coll} = n \sigma v = 1/\tau_{0}$ is the collision width.

If resonance is placed in a medium, it acquires an additional broadening.
Its spectral function modifies accordingly, as shown schematically on Fig. 1.

{\it Collision broadening modifies spectral functions of resonances.}

These ideas are in agreement with the uncertainty relation:
\begin{equation} 
\Delta E \ge \frac{1}{\tau_{0}} = \frac{v}{\ell_f} = n \sigma v = \Gamma_{coll}. 
\label{uncert}
\end{equation}
Experimentally it is not possible to resolve transition energy 
between two atomic levels with accuracy $\Delta E$ better than the inverse 
collision time $\tau_{0}$. If we would ask an experimentalist to construct 
an experimental distribution of photon energies (wavelengths) e.g. in the 
Balmer $\beta$-line, he could present something resembling the curves on Fig. 1. 

\subsection{Experimental observations of modified profiles of atomic spectral lines}

Shifts of energy levels and collision broadenings of atoms are observed experimentally 
in gases. Let us provide a couple of illustrations. Fig. 2 (left) recalls the structure of 
the hydrogen atom energy levels and the associated transitions and 
shows (right) the {\it modified} spectral density of the $H_{\beta}$ line of 
a hydrogen-like atom in a gas. 
\vspace{-5mm}
\begin{figure}[!htb]
\begin{center}
\includegraphics[angle=0,width=10cm]{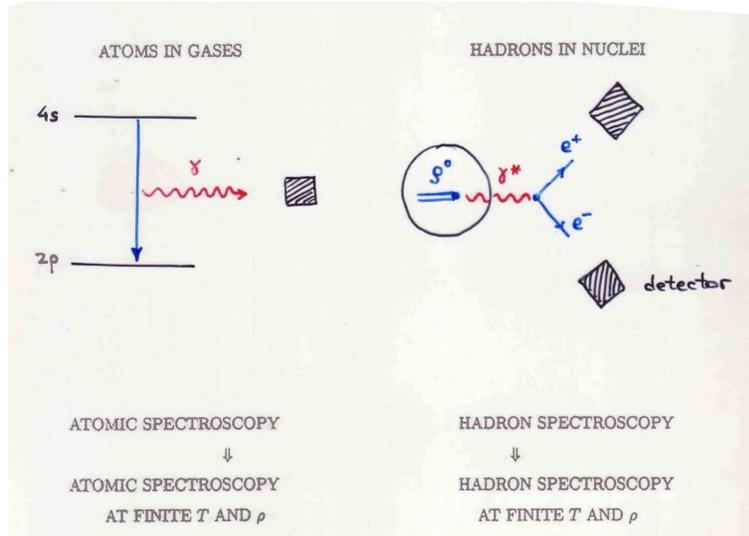}
\caption{Remarkable analogy between atomic spectroscopy and hadron 
spectroscopy.}
\label{sig11_figu}
\end{center}
\end{figure}
\vspace{-5mm}

\section{Nucleon resonances in nuclei}


After clarifying principal issues of spectroscopy of isolated atoms and molecules 
(as a result of which quantum mechanics appearred), atomic physics evolved towards 
studying the in-medium modifications of atoms at finite 
temperature and density. Physics of intermediate energies $\sim$ 1 GeV follows in 
its evolution the general trend of atomic physics with a time lag $\sim$ 70 years. 
Last 15 years, new possibilities occurred to accomplish experimentally similar program for hadron 
spectroscopy. This analogy is illustrated in Fig. \ref{sig11_figu}.

In nuclear physics, an example of broadening of resonance profiles is 
delivered by experiments on photoabsorption on nuclei \cite{Bianchi:1993nh}:
\vspace{-3mm}
\begin{figure}[!htb]
\begin{center}
\includegraphics[angle=0,width=4cm]{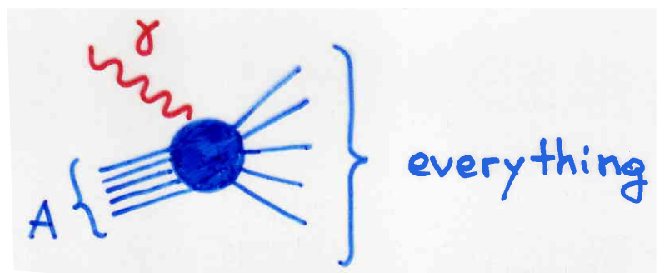}
\end{center}
\end{figure}
\vspace{-5mm}

\begin{figure}[!htb]
\begin{center}
\includegraphics[angle=0,width=6cm]{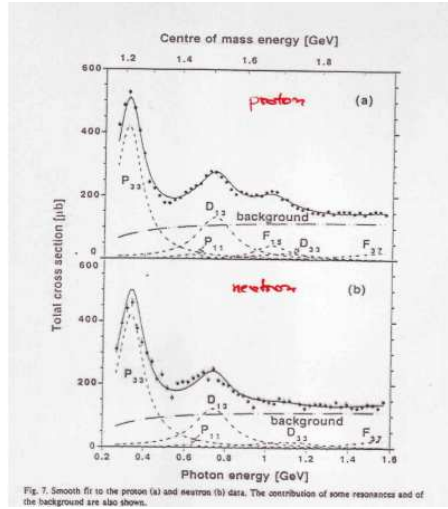}
\caption{Total photoabsorption cross section on proton (a) and neutron (b)
vs photon energy.
Contributions of various nucleon resonances and the background are shown
(from Ref.\cite{Kondratyuk:1994ah}).
}
\label{figure1}
\end{center}
\end{figure}

The photoabsorption cross sections on free nucleons ($A = 1$) develop clear
peaks associated to various nucleon resonances $R = \Delta(1232),\;N^{*}(1440),\ldots$ 
(see Fig. \ref{figure1}). 
The resonance widths are related physically to $R \to N\pi,\;N\pi\pi$ decays
e.g. to radiation damping of Sect. 2.1.

\vspace{-4mm}

\begin{figure}[!htb]
\begin{center}
\includegraphics[angle=0,width=6cm]{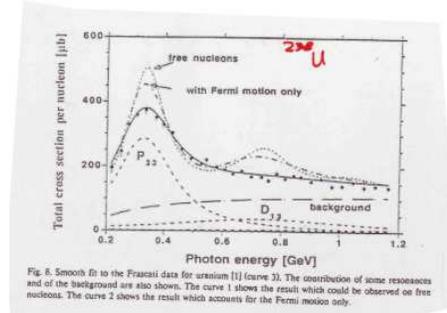}
\caption{Total photoabsorption cross section on $^{238}$U versus photon energy
in the laboratory system.
Contributions of various nucleon resonances $R = \Delta(1232),N^{*}(1440),\ldots$ and the background (long-dashed curve) are shown. The 
additional broadening is related to the Doppler effect (Fermi motion) and 
collision broadening. The Pauli blocking effect is included. The 
modification of the in-medium spectral density of resonances is clearly seen. 
Resonances heavier than $\Delta(1232)$ are broadened and masked by the 
background (from Ref.\cite{Kondratyuk:1994ah}).
}
\label{figure2}
\end{center}
\end{figure}

In nuclear environment, the Doppler effect (Sect. 2.2) related to the Fermi motion
of nucleons and collision broadening (Sect. 2.3) come into play. The result
is presented on Fig. \ref{figure2}. 

\section{In-medium properties of vector mesons: Theoretical models}

The change of the nucleon mass in nuclear matter was discussed first in
the pioneering works by Walecka and Chin \cite{Walecka:1974qa,Chin:1977iz}
already in the 1970's. The Mean Field and Relativistic Mean Field (RMF)
approximations were developed to treat the dense matter self-consistently to
all orders in the density. The effective nucleon mass was found to decrease
significantly in nuclei at saturation density. This effect predicted
by phenomenological models is confirmed by the developments
of QCD. 

During the last decade, the problem of the description of hadrons in dense
and hot nuclear matter received new attention due to the possibility to test
theoretical predictions with  heavy-ion experiments. This
problem is related to chiral symmetry breaking and QCD confinement. 
Being intrinsically non-perturbative, the number of available
theoretical schemes is limited. 
The knowledge of the non-perturbative QCD comes in particular from 
lattice simulations. However, lattice results suffer still from 
uncertainties due to finite lattice size effects and 
the proper inclusion of fermions. 
An important role is played by field theoretical  models which
incorporate fundamental features of QCD.

The model proposed by Walecka \cite{Walecka:1974qa} was further analyzed in
Refs. \cite{Chin:1977iz,BGP,Mishra} to study the in-medium vector mesons. 
The Nambu-Jona-Lasinio (NJL) model was introduced in the sixties as a theory
of interacting nucleons \cite{nambu} and later it was reformulated in terms
of quark degrees of freedom. This model was used by the T\"{u}bingen group 
\cite{Tsushima:1991fe} to study matter under the extreme conditions. The
movement towards the chiral symmetry restoration is reliably described
within the NJL model \cite{BR2004}. 
QCD sum rules provide a successful analytic tool to work in the
non-perturbative regime. The medium modifications of hadrons are discussed
in QCD sum rules at finite temperature and density \cite
{Boch,Drukarev:1988ib,Adami:1993tp,QCDSR,KKW}. The vector self-energy and
the effective nucleon mass at saturation density agree with those
obtained by the conventional methods of nuclear physics. The values of QCD
condensates and the nucleon expectation values of the various quark
operators including the chiral order parameter can be determined within this
approach \cite{DRUK,Chanfray}.

Dispersion theory combined with the optical potential method has been
used by Weise et al. \cite{KKW,KS,Martell2004} to calculate mass
shifts, broadening and spectral densities for vector mesons in the
presence of nucleons and pions at finite density and temperature. The
experimental data on the total cross sections are used to saturate the
scattering amplitude at low energies with resonances. However, 
the obtained results are at variance with \cite{BR}. 
Unitarized thermal chiral perturbation theory has been used in 
\cite{UNCHPT,Boris}. There one matches 
unitarized amplitudes smoothly with the ChPT loop expansion in
the low-energy region. The estimates show, in particular, mass shifts and a
clear increase of the thermal widths of unflavored mesons which have 
relevance within the context of ultra-relativistic heavy-ion
collisions. 
Other attractive phenomenological models for the in-medium unflavored and
charmed mesons have also been proposed in \cite{BR,Micro,weise99,digal,brat2}. 

The current status of the Brown-Rho scaling \cite{BR} has recently been discussed by
Brown and Rho \cite{BR2004}.

\section{In-medium properties of vector mesons: Experiments on dilepton production}

The available experiments on the dilepton production are summarized in Table 1.

\begin{table}
\tbl{The enhanced production of dileptons in low and intermediate mass continuum. 
The fifth column shows intervals of invariant masses of dileptons. 
$O/E$ is the ratio between 
integral numbers of observed and expected dileptons.}
{\begin{tabular}{@{}lllcll@{}}\toprule
Collaboration & Ref. &E/A   & Dileptons              & M    & O/E      \\ 
              &      &[GeV] &                        & [GeV]&          \\
\colrule
NA38 at CERN  & \cite{NA38}  &200 - 450    & ${\mu }^{+}{\mu }^{-}$ & 0.6 - 6   & $\sim 1.5$  \\ 
CERES         & \cite{ceres} &200 - 450    & ${e   }^{+}{ e  }^{-}$ & 0.05 - 1.4  & $5 \pm 3$ \\ 
HELIOS-3      & \cite{HELIOS}&200 - 450    & ${\mu }^{+}{\mu }^{-}$ & 0.3 - 4   & $\sim 2.5$\\ 
DLS at BEVALAC& \cite{DLS}   &1 - 5    & ${e   }^{+}{e   }^{-}$ & 0.05 - 1  & $2-3$     \\ 
KEK           & \cite{KEK}   &12         & ${e }^{+}{e }^{-}$     & 0.05 - 2  & $\sim 2$  \\ 
NA60          & \cite{NA60}  &160        & ${\mu }^{+}{\mu }^{-}$ & 0.2 - 1.2  & $\sim 2$     \\ 
HADES at GSI  & \cite{HADES} &1 - 2       & ${e    }^{+}{e }^{-}$ & 0.05 - 1  & $\sim 1$\\ 
\botrule
\end{tabular}
}
\label{aba:tbl1}
\end{table}

The dilepton spectra measured by the CERES \cite{ceres} and HELIOS-3 \cite{HELIOS} 
Collaborations at CERN SPS found a significant enhancement of the low-energy dilepton yield below the $%
\rho $ and $\omega $ peaks \cite{ceres} in heavy systems ($Pb+Au$) as
compared to light systems ($S+W$) and proton induced reactions ($p+Be$).
Theoretically, this enhancement can be explained assuming a dropping mass
scenario for the $\rho $ meson and the inclusion of in-medium spectral
functions for the vector mesons \cite{rapp,cassing99}. The enhanced low
energetic dilepton yield originates to most extent from an enhanced
contribution of the $\pi ^{+}\pi ^{-}$ annihilation channel. An alternative
scenario is the formation of a quark-gluon plasma in the heavy systems which
leads to additional contributions to the dilepton spectrum from perturbative
QCD ($p$QCD) such as quark-antiquark annihilation or gluon-gluon scattering 
\cite{rapp,weise00}.

The dilepton mass spectrum measured at KEK in p + A reactions at the beam
energy of 12 GeV \cite{KEK} revealed an excess of the dileptons below the $%
\rho $-meson peak over the known sources. These data were analyzed in Ref. 
\cite{EB} with no success to reproduce the experimental spectrum within a
dropping mass scenario's and/or a significant collision broadening of the
vector mesons. 

The results of the DLS \cite{DLS}, NA60 \cite{NA60}, and HADES \cite{HADES}
collaborations are discussed in Sect. 9.

\section{Elementary sources of dileptons}

The direct decay mode $\rho,~\omega \to e^{+}e^{-}$ represents a signal. There 
are many other 
sources, however, which constitute a background and which should accurately 
be subtracted.

\subsection{Dilepton modes in decays of light unflavored mesons}

One has to distinguish between vector, pseudoscalar and scalar mesons ${\mathcal M} = V,\;P,\;S$, 
respectively, where 
\begin{equation} 
V = \rho,~\omega,~\phi;~~~~P = \pi,~\eta,~\eta^{\prime};~~~~S = f_{0}(980),~a_{0}(980). \nonumber 
\end{equation} 

Decay modes are presented below:
\newpage
\begin{table}
\begin{center}
\begin{tabular}{llll}
$V \to e^{+}e^{-}$ & & & Direct decays \\ 
$P \to \gamma e^{+}e^{-}$ & $S \to \gamma e^{+}e^{-}$ & & Dalitz decays \\
$V \to P e^{+}e^{-}$ & $P \to V e^{+}e^{-}$ & & Dalitz decays \\
$V \to PP e^{+}e^{-}$&$P \to PP e^{+}e^{-}$&$S \to PP e^{+}e^{-}$ & Four-body decays\\
\end{tabular}
\end{center}
\end{table} \vspace{-6mm}

The photon and dilepton branching ratios are calculated using the 
effective meson theory \cite{FFKR}. Mesons are produced in $NN$ and $\pi N$
collisions. The reactions of rescattering, absorption and reabsorption of mesons
affect the observed dilepton yield.

\subsection{Dilepton modes in decays of nucleon resonances}

Besides light unflavored mesons ${\mathcal M} = V,\;P,\;S$, nucleon 
resonances $R = \Delta(1232),\;N^{*}(1440),\ldots$ are produced in the course of heavy-ion collisions. Their decays contribute to the measured dilepton spectra.

The $\Delta(1232)$ Dalitz decay is one of the major sources of dileptons in heavy-ion 
collisions at intermediate energies \cite{FRA,GI,TUE}. The first correct calculation of that decay is given only recently \cite{krivo02}, while kinematically complete expressions for Dalitz decays of other positive and negative parity high-spin resonances are given in Ref. \cite{AOP}. 

In order to evaluate the Delta Dalitz decay rate, HADES \cite{HADES} used  one of six pairwise different incorrect expressions available in literature before \cite{krivo02}. We present therefore results of Ref. \cite{krivo02}: 

The $\Delta $ resonance width for decay into nucleon and a virtual photon is given by
\begin{eqnarray}
\Gamma (\Delta  &\rightarrow &N\gamma ^{*})=\frac{\alpha }{16}\frac{%
(m_{\Delta }+m_{N})^{2}}{m_{\Delta }^{3}m_{N}^{2}}((m_{\Delta
}+m_{N})^{2}-M^{2})^{1/2}  \nonumber \\
&&((m_{\Delta }-m_{N})^{2}-M^{2})^{3/2}\left( G_{M}^{2}+3G_{E}^{2}+\frac{%
M^{2}}{2m_{\Delta }^{2}}G_{C}^{2}\right) .  \label{width_D}
\end{eqnarray}
Here, $m_{N}$ and $m_{\Delta }$ are nucleon and $\Delta $ masses, 
$M^{2}=q^{2}$ is the photon four-momentum, $G_{M}$, $G_{E},$ and $G_{C}$ 
are magnetic, electric and Coulomb transition form factors.
 
The factorization prescription allows to find the dilepton decay rate of 
the $\Delta $ resonance:
\begin{equation}
d\Gamma (\Delta \rightarrow Ne^{+}e^{-})=\Gamma (\Delta \rightarrow N\gamma
^{*})M\Gamma (\gamma ^{*}\rightarrow e^{+}e^{-})\frac{dM^{2}}{\pi M^{4}},
\end{equation}
with 
\begin{equation}
M\Gamma (\gamma ^{*}\rightarrow e^{+}e^{-})=\frac{\alpha }{3}%
(M^{2}+2m_{e}^{2})\sqrt{1-\frac{4m_{e}^{2}}{M^{2}}}
\label{fina}
\end{equation}
being the decay width of a virtual photon into the dilepton pair with
invariant mass $M$. The last three equations being combined give the $\Delta
(1232)\rightarrow Ne^{+}e^{-}$ decay rate.

The available experimental data on electro- and photoproduction experiments 
with $\Delta(1232)$ are fitted in Ref. \cite{AOP} using the eVMD framework to give
\begin{eqnarray}
G_{M} &=& (2.461 - 0.485 M^2 - 0.004 M^4)G,\label{param1} \\
G_{E} &=& (0.062 - 0.010 M^2 + 0.004 M^4)G,\label{param2} \\
G_{C} &=& (0.518 - 0.087 M^2)G             \label{param3}
\end{eqnarray}
where $M^2$ is in GeV$^2$ and
\begin{equation}
G = \prod_{i=1}^{4}\frac{m_{i}^2}{m_{i}^2 - im_{i}\Gamma_{i}(M) - M^2}.
\label{GD}
\end{equation}
The masses $m_{i}$ of the $\rho$-meson family members are the following: 
0.769, 1.250, 1.450, and 1.720 GeV. The $\rho$-meson decay rate $\Gamma_{1}(M)$
is proportional to two-pion phase space and normalized to the vacuum width, 
whereas $\Gamma_{i}(M)$ for $i \geq 2$ are set equal to zero, since $M \leq 1$ GeV 
at our conditions. The quality of the fit can be verified with Fig. 20 of Ref. \cite{AOP}. 
Equations (\ref{param1}) - (\ref{GD}) are in agreement with the quark counting rules, the condition $G_{M}/G_{E} \to - 1$ at $M^2 \to - \infty$ is a consequence of the quark counting rules.

\section{Microscopic eVMD model}

Radially excited $\rho $- and $\omega $-mesons are introduced \cite{resdec,AOP} 
to ensure the correct asymptotic behavior of the $RN\gamma $ transition 
form factors in line with the quark counting rules \cite{Matveev:1973ra}. 
We refer this model as the extended vector meson dominance (eVMD) model. 
From the experimental side, the excited vector mesons are needed to match
photon $M^2 = 0$ and vector meson $M^2 \sim 0.5$ GeV$^2$ experimental branchings 
of nucleon resonances within a unified scheme.

\begin{itemize}
\item  The eVMD model provides a unified description of the photo- and
electro-production data and of the vector meson and dilepton decays of
nucleon resonances and accounts for quark counting rules.
\end{itemize}

We take constraints on the transition form factors from the quark counting rules
into account explicitly. The remaining parameters are
fixed by fitting the available photo- and electro-production data and using
results of the  $\pi N$ multichannel partial-wave analysis. When data are not 
available, we used predictions of the additive quark models.

Radially excited $\rho $- and $\omega$-mesons as a consequence of the quark 
counting rules interfere with the ground-state mesons {\it destructively} below 
the $\rho$-meson peak and reduce dilepton spectra from decays
of nucleon resonances in the vacuum, accordingly.

\section{How to keep photon massless with $\rho^{0}-\omega-\gamma$ mixing?}

VMD model and its modifications introduce mixing of photon with 
vector mesons $\rho^{0}$, $\omega$, $\phi$, etc. Such a mixing can, in principle, 
generate finite photon mass and destroy gauge invariance. This problem has been 
solved for VMD model by Kroll, Lee and Zumino \cite{KROLL} by constructing 
an effective Lagrangian for photons and vector mesons, which reproduces VMD 
predictions at a tree level. We describe a distinct consistency proof. 

We start from an effective Lagrangian involving pions interacting with 
photons. An example of such a Lagrangian is e.g. the non-linear sigma model. The
vector mesons appear as resonances in two-pion scattering channel ($\rho$-mesons)
and three-meson scattering channel ($\omega$-mesons). 


\begin{figure}[!htb]
\begin{center}
\includegraphics[angle=0,width=5cm]{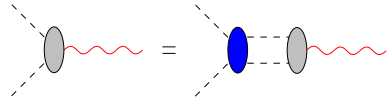}
\caption{
Diagram representation of the FSI of pions (dashed lines) contributing to
form factor in the $\rho$-meson channel. The wavy lines are photon lines.}
\label{fig1}
\end{center}
\end{figure}

Let us consider an absorption of a photon in an isovector channel shown on Fig. \ref{fig1}.
Applying two-body unitarity and taking into account analyticity (see e.g. \cite{BD}, Chap. 18), we replace the pointlike vertex $e$ by $eP_{l}(t)/D_{J}(t)$ where $t=q^2$ is the photon momentum squared, $P_{l}(t)$ is a polynomial of the degree $l$, and $D_{J}(t)$ is the Jost function defined in terms of the $p$-wave isovector two-pion scattering phase shift $\delta(t)$:
\begin{equation}
D_{J}(t) = \exp(-\frac{t}{\pi}\int_{t_{0}}^{\infty}\frac{\delta(t^{\prime})dt^{\prime}}{t^{\prime}(t - t^{\prime})})
\label{JOST}
\end{equation}
where $t_{0}$ is the two-pion threshold.

In the no-width approximation, the phase shift accounting for the existence of $n$ resonances is given by
\begin{equation}
\delta(t) = \sum_{i=1}^{n}\pi \theta(t - m^2_{i})
\label{J2}
\end{equation}
where $m_{i}$ is the mass of the $i$-th radial excitation of the $\rho^{0}$-meson. Substituting this expression into Eq.(\ref{JOST}), we obtain
\begin{equation}
F(t) = P_{l}(t)\prod_{i=1}^{n}\frac{m^2_{i}}{t - m^2_{i}}
\label{J3}
\end{equation}
(cf. Eqs.(\ref{param1}) - (\ref{GD})). The requirement $F(t) \rightarrow 0$ at $t \rightarrow \infty$ gives $l<n$. 

Analytical functions are fixed by their singularities. The representation (\ref{J3}) can be rewritten in the additive form
\begin{eqnarray}
F(t) = \sum_{i=1}^{n}c^{i}\frac{m^2_{i}}{m^2_{i} - t}
\label{DE3}
\end{eqnarray}
where $c^{i}$ are some coefficients. The normalization condition $F(0) = 1$ and quark counting rules impose constraints (sum rules) for $c^{i}$. 

The effective pion Lagrangian is well defined, since pions are stable particles which exist as asymptotic states. In the approach presented above, the problem of gauge invariance does not appear, since gauge invariance of the effective Lagrangian ensures transverse polarization tensor of photons and the vanishing photon mass. The vector mesons
are resonances accounted for by the the final-state interactions (FSI). 

\section{In-medium properties of vector mesons: What can we learn from observations?}

The gap between observables measured in heavy-ion collisions and theoretical
models of the in-medium hadrons is filled by transport models which account
for complicated dynamics of heavy-ion collisions and provide a link between
theory and experiment. We comment results concerning physical properties 
of in-medium vector mesons that can be derived from experimental data with 
use of transport models.

\subsection{A constraint to $\omega$-meson collision width from DLS and HADES data}

In-medium broadening of vector mesons gives an increase of the nucleon resonance 
decay widths $R \rightarrow NV$ and a decrease of the dilepton branchings 
$V \rightarrow e^+ e^-$  due to enhanced total vector meson widths. 

\begin{figure}[!htb]
\begin{center}
\includegraphics[angle=0,width=6cm]{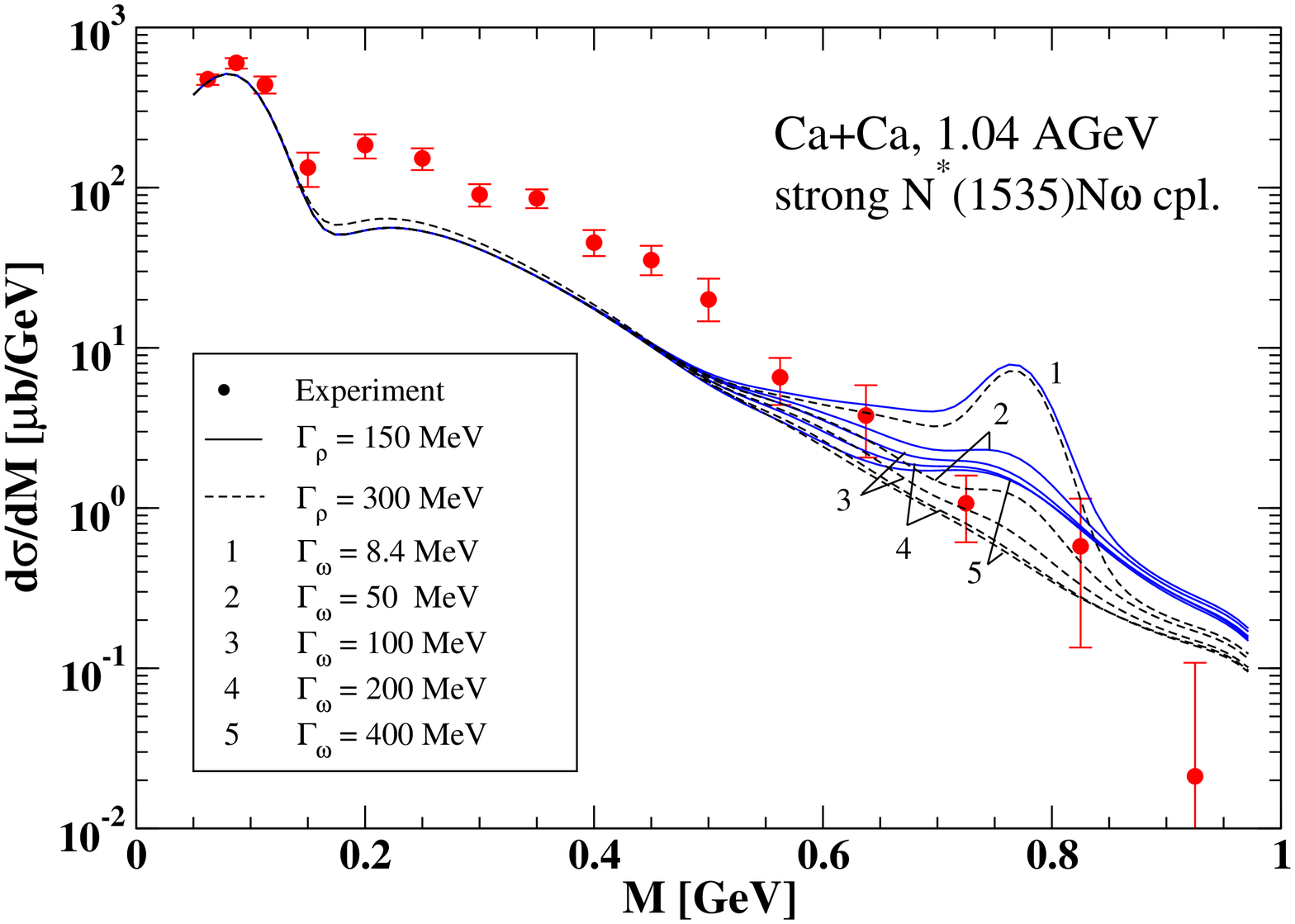}
\end{center}
\caption{Dilepton spectra in $Ca + Ca$ collisions for different 
values of the in-medium $\rho$ and $\omega$ widths.
The solid curves correspond calculations where the $\rho$ width is 
kept at its vacuum value of $150$ MeV (no collision broadening). 
The dashed curves correspond to a total $\rho$ width of 300 MeV. 
In both cases the $\omega$ width is varied between  
$\Gamma_{\omega}^{tot} = 8.4\div 400$ MeV. 
$\Gamma_{\omega}^{tot} \geq 50$ MeV holds as a conservative constraint (from \cite{TUE}).
}
\label{DLS_fig3}
\end{figure}

The differential branching $dB(\mu,M)^{R\rightarrow NV}$ of resonance $R$ with 
an off-shell mass $\mu$ increases with the $V$ meson width  
due to subthreshold character of the vector meson production in
light nucleon resonances. The dilepton branching of the nucleon resonances 
\begin{equation}
dB(\mu)^{R\rightarrow Ne^+e^-} \sim
dB(\mu)^{R\rightarrow NV} \frac{\Gamma_{V \rightarrow e^+e^-}}{\Gamma_V^{\rm total}}
\label{bra1}
\end{equation}
is, on the other hand, inverse proportional to the {\it total} vector 
meson width. This effect is particularly strong for $\omega$, since 
$\Gamma_{\omega}^{vac} = 8$ MeV only. Fig. \ref{DLS_fig3} shows the collision broadening 
effect on the dilepton spectra. To get description of the data, we have to assume
\begin{equation}
\Gamma_{\omega}^{total} \approx \Gamma_{\omega}^{coll} \ge 50 ~{\rm MeV}~~~{\rm at ~density}~~~\rho \sim 1.5\rho_{0}
\label{constr}
\end{equation} 
where $\rho_{0}$ is the saturation density. The similar conclusions holds for HADES data \cite{HADES}.

\subsection{Evidence for decoherence in transition form factors of nucleon resonances 
from DLS and HADES data}

\begin{figure}[!htb]
\begin{center}
\includegraphics[angle=0,width=5cm]{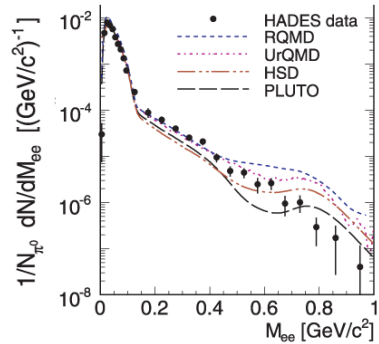}
\caption{
  The experimental dilepton spectrum as compared to predictions of PLUTO \cite{HADES} thermal 
model, UrQMD \cite{FRA}, RQMD \cite{TUE}, and HSD \cite{cassing99} transport 
models. 
}
\label{HADES}
\end{center}
\end{figure}

Fig. \ref{DLS_fig3} shows an obvious  deficit of dileptons below the $\rho$-meson peak. This phenomenon has been called "DLS puzzle", since transport models were not able to reproduce it. The eVMD suggests a medium-induced decoherence of vector mesons entering transition form factors of nucleon resonances and gives an enhancement of dilepton yield \cite{TUE}. It does not remove disagreement with the DLS data completely, however, the recent HADES data \cite{HADES}
shown on Fig. \ref{HADES} are reproduced at $M \leq m_{\rho}$ perfectly \cite{Cozma, HADES}. 
A more detailed theoretical study of the decoherence effect would certainly be welcome.

\subsection{No evidence for dropping $\rho$-meson mass with increasing 
density from NA60 data}

The NA60 Collaboration presented recently an impressive attempt to extract 
the $\rho$-meson spectral function from dilepton spectra at ultrarelativistic 
heavy-ion collisions \cite{NA60}. It had not found evidence for a change 
of the $\rho$-meson mass. This observation 
poses obvious difficulties for the Brown-Rho scaling hypothesis.

\section{Conclusions}

In this lecture, an introduction is made to physics of in-medium behavior 
of resonances. Transparent methods developed in the atomic spectroscopy in gases 
can be useful for in-medium hadron spectroscopy. 
The main topic we focused on is the modifications of spectral 
functions. Starting from instructive examples in atomic and nuclear physics, 
we approached the problem of vector mesons description in dense nuclear matter. 

Theoretical models are discussed in Sect. 4. The most of them do not go beyond the 
first order in density. In the non-linear sigma model, density expansion for pions has zero 
convergence radius \cite{DENS}. The higher-order calculations would help to clarify 
to what extent density expansion for vector mesons is reliable.

This summer (2006 year) two new important experimental works \cite{NA60,HADES} from NA60 and HADES 
collaborations have been completed and results published. Preliminary conclusions 
are provided in Sect. 9. 
\begin{itemize}
\item Collision broadening of $\omega$-meson is constrained from below by $\sim 50$ MeV at $\rho \sim 1.5 \rho_{0}$ where $\rho_{0}$ is the saturation density.  
\item DLS puzzle had apparently dissolved due to decoherence in in-medium propagation of vector mesons.
\item The NA60 data \cite{NA60} do not provide any evidence for dropping the $\rho$-meson mass.
\end{itemize} 

Among difficulties in description and interpretation of the HADES data, one has to mention theoretical excess of dileptons above the $\rho$-peak. The dilepton yield 
could be suppressed by increasing the $\rho$ and $\omega$ collision widths and by taking into account collision broadening of nucleon resonances.  

The results of the KEK Collaboration \cite{KEK} show an enhancement below the $\rho$-peak, which is not understood.

Results of the dilepton production in heavy-ion collisions attract great attention and will certainly be analyzed in future. 

\bibliographystyle{ws-procs9x6}
\bibliography{ws-pro-sample}

\end{document}